\documentclass[10pt,conference]{IEEEtran}
\usepackage{epsfig,setspace,amsmath,epsf,amssymb,bm,theorem,cite,graphicx,epstopdf,algorithm,algpseudocode,float,color,mathtools,authblk,tikz,physics}
\usepackage{booktabs}
\usepackage{subfigure}
\usepackage{bbm}
\usepackage[short,c2]{optidef}

\IEEEoverridecommandlockouts
\allowdisplaybreaks

\begin{document}

\title{When and Which Sensor to Observe? \\ Timely Tracking of a Joint Markov Source

\thanks{I. Cosandal and S. Ulukus are with the University of Maryland, College Park, MD, USA. N. Akar is with Bilkent University, Ankara, T\"{u}rkiye. Corresponding author: S. Ulukus (email: ulukus@umd.edu).

This work is done when N.~Akar was on sabbatical leave as a visiting professor at the University of Maryland, MD, USA.}}


\author{Ismail Cosandal \qquad Sennur Ulukus \qquad Nail Akar}

\maketitle

\begin{abstract}
We investigate the problem of remote estimation (at a monitor) of a discrete-time joint Markov process with individual components which can be observed with dedicated sensors. At a given time slot, the monitor has the option of staying idle or sending a pull request to one of the sensors to obtain a partial state value, while the sensors are assumed to have heterogeneous sampling costs. Our goal is to develop a monitor pull policy, i.e., determining \emph{when} and \emph{towards which} sensor to send a pull request, in order to minimize a weighted sum of average age of incorrect information (AoII), or in short \emph{age}, and sampling costs. As the communication model, we assume an erasure channel with a fixed one-slot delay from each sensor to the monitor. In this setting, the monitor does not perfectly know either the state of the process or the age, at any given time. We first obtain a sufficient statistic, namely \emph{belief}, representing the joint distribution of the age and the current state of the observed process, by using the history of all pull requests and observations. Then, we formulate the optimization problem as a continuous state-space Markov decision process (MDP), namely \emph{belief-MDP}, for the solution of which we propose two model predictive control (MPC) methods, namely MPC without terminal costs (MPC-WTC), and reinforcement learning MPC (RL-MPC). The effectiveness of the proposed methods is validated by numerical examples.
\end{abstract}


\section{Introduction}
Age of incorrect information (AoII) is a joint mismatch and freshness metric that captures for how long there has been a mismatch between an observed random process and its estimate at a remote monitor \cite{maatouk2022age}. AoII takes into account the dynamics of the source process, and is thus considered to be a semantic metric \cite{lu2022semantics, maatouk2022age}. Different from other mismatch metrics such as the mean squared error (MSE) or binary freshness (BF) \cite{bastopcu2021,cosandal2023timely}, AoII penalizes the duration of incorrect estimation by increasing linearly with time, and the AoII process is reset to zero as soon as the monitor correctly estimates the process. AoII is fundamentally different from other information freshness metrics derived from the age of information (AoI)  \cite{bastopcu2020should,ayan2019value}, since the AoI process can only drop upon the reception of a status update packet, while AoII can be reset not necessarily only with a status update, but with an update of the estimate of the monitor, or upon a state transition at the source to the estimation value at the monitor.

\begin{figure}[t]
    \centering
    \includegraphics[width=0.95\linewidth]{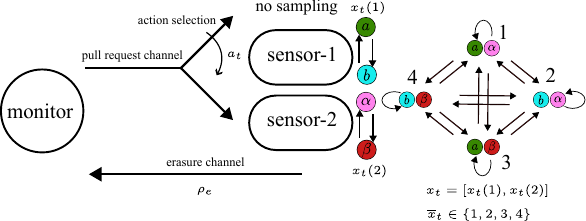}
    \caption{The illustration of the pull-based status update system. The two processes that sensors observe may not be individually Markov, but their joint process is Markov with its transition probabilities known to the monitor. State-space of the joint process $x(t)$ is ordered as $\{(a,\alpha),(b,\alpha),(a,\beta),(b,\beta)\}$ with the index process $\overline{x}_t$ taking the index of the joint state, as its value.}
    \label{fig:sys}
    \vspace*{-0.4cm}
\end{figure}

In this paper, we consider a unique scenario involving multiple sensors each of which observing a single component (i.e., partial state) of a joint Markov process, and a monitor which aims to minimize a weighted combination of the time-averaged AoII (equivalently, the mean of the steady-state AoII, denoted by MAoII) and sampling costs, by either staying idle or selecting one of the sensors to observe a partial state, at each time slot. For the communication model, we assume that pull requests reach the sensors instantaneously on the reverse channel from the monitor towards the sensors. However, for the forward channel from the sensors towards the monitor, we assume that transmissions take place in one-slot with a given erasure probability. 

As a simple motivating example, consider an object performing a random walk on a $\text{2D}$ grid, where two sensors observe the object's $x$ and $y$ components separately, and a monitor remotely track the location of the object. In this example, the monitor should make decisions on when and which sensor to observe, by considering the system dynamics and all previous observations, to track the object's location, under energy consumption constraints. As another example, consider a joint fire/temperature process which is tracked by a remote monitor using two sensors; one sensor observing the temperature level (as low, moderate, high), and the other sensor observing the event of a fire (as fire, no fire). In this example, we assume that the fire event only occurs if the temperature is high, thus the individual events may not be Markov processes individually, but the joint process can be assumed to be Markov. 

As a result of the communication model with delays, partial observability, and errors in the forward channel, the monitor can never be exactly sure of the exact state of the process, and hence, the AoII value of its estimate. However, it is possible for the monitor to obtain the so-called \emph{belief} which corresponds to the joint distribution of the state of the Markov process and the AoII \cite{cosandal2024joint} based on all previous observations and actions. In addition, the monitor estimates the original process using the maximum a-posteriori (MAP) rule, which is also a function of the belief. This estimation method affects the distribution of the AoII, thus the transition probabilities among the beliefs and the cost of a belief. This relation hinders us from using the partially observable MDP (POMDP) formulation directly \cite{Krishnamurthy_2016}. On the other hand,  alternative formulations are proposed as extensions of the POMDP framework for problems where the cost or the state transition probabilities depend on the belief \cite{araya2010pomdp, spaan2015decision, satsangi2018exploiting}. These extensions of POMDPs \cite{fehr2018rho}, and the POMDP itself \cite{eck2016potential}, can be expressed with an equivalent continuous-state MDP where states are the distributions of unobserved states, which gives rise to a formulation called the \emph{belief-MDP}.

We will now overview the existing works on the solution of belief-MDPs which are most relevant to the current paper. The state-space of the belief-MDP, which is termed as \emph{belief space}, contains uncountably many elements, and therefore it is not possible to obtain the value function for each belief state by using dynamic programming from the belief-MDP representation. One method to cope with the continuous belief space is the discretization of the belief space \cite{cregg2024reinforcement} which however suffers from low accuracy and high complexity for high and low discretization step sizes, respectively \cite{hauskrecht2003linear}. In some cases \cite{liu2023optimizing, liu2023optimizing2}, only a finite number of beliefs can be visited; hence, it is possible to convert the belief-MDP to an MDP with a finite state-space, which does not apply to the problem of interest here. Although an uncountable number of belief values can be visited in the current paper's setting, the number of possible next beliefs from an individual belief in a finite horizon is countable due to the finite-dimensional observation and action spaces. In this paper, we utilize model predictive control (MPC) which evaluates all possible outcomes for a finite number of steps and adds terminal costs of states in the final step \cite{bertsekas2012dynamic}. Subsequently, MPC obtains the action sequence that minimizes the expected cost from all calculated trajectories, and applies the first action from this sequence. This process is then repeated at each slot. MPC has successfully been used to solve belief-MDPs in several recent works owing to its ability to decide based on future costs in a finite horizon \cite{sehr2018performance, ulfsjoo2022integrating, esfahani2021reinforcement}.

Regarding terminal costs, we consider two approaches. The first approach is MPC without terminal cost (MPC-WTC) which considers all terminal costs to be equal (zero in particular), and the first action of the path which minimizes the expected cost on a finite horizon is taken \cite{boccia2014stability, grune2012nmpc}. The main advantage of this method is ease of implementation since it does not require offline learning. However, a long time horizon may be required for improving performance at the expense of increased complexity of evaluating all trajectories. The second approach is based on reinforcement learning MPC (RL-MPC) where terminal costs are approximated with the aid of RL \cite{lin2023reinforcement}. RL-MPC utilizes MPC-WTC to learn terminal costs in the first iteration, and increases its approximation horizon iteratively in a similar fashion to fitted Q-learning \cite{ernst2003iteratively}. This method allows us to find an action sequence that minimizes the cost for a larger horizon with a lower complexity compared to MPC-WTC. On the other hand, approximation errors encountered at previous iterations may lead to performance degradation \cite{lin2024learning}, and offline learning is required before run-time as opposed to MPC-WTC. 

The main contributions of our paper can be summarized as follows:
\begin{itemize}
    \item We investigate a timely tracking problem where a joint Markov process is observed partially only and with delays, at an observation instant, which to the best of our knowledge, has not been explored before, in the literature.
    \item Considering heterogeneous sampling costs for the sensors, we aim to obtain a monitor pull policy which minimizes the weighted sum of MAoII and overall sampling costs. For this purpose, we adopt the method for calculating the joint distribution of age and state in \cite{cosandal2024joint} for the case of partial observations, and formulate the optimization problem as a belief-MDP. 
    \item  The proposed belief-MDP has a continuous state-space for which the use of conventional dynamic programming techniques is infeasible. To solve the belief-MDP problem, we propose two MPC methods, namely MPC-WTC and RL-MPC, and validate the effectiveness of the two proposed approaches, in terms of performance and computational complexity, through extensive numerical experimentation.
\end{itemize}

The remainder of the paper is organized as follows. In Section~\ref{sec:RW}, related work from the literature is presented. The system model is given in Section~\ref{sec:SM}. The optimization problem is formulated as a belief-MDP in Section~\ref{sec:BMDP}, and Section~\ref{sec:MPC} is dedicated to the proposed MPC solutions. Numerical results are presented in Section~\ref{sec:NR}, and the conclusions are given in Section~\ref{sec:conc}. 

\section{Related Work} \label{sec:RW}
POMDP is an extension of MDP \cite{Krishnamurthy_2016} where the action taker cannot directly observe the state of the problem, but it can estimate the likelihood of the state from partial observations, which is called the \emph{belief}, and the policy is defined as a function of the belief \cite{smallwood1973optimal, spaan2012partially}. In this formulation, the underlying process governing state transitions is a Markov process, and state transition probabilities and the cost function depend only on the state and the action. The study \cite{araya2010pomdp} further extends POMDP to address problems of belief-dependent cost, namely $\rho$POMDP, and demonstrates that if the cost function is convex, existing POMDP solutions can be adapted for $\rho$POMDP. In addition, the study \cite{spaan2015decision} considers a scenario in which an agent aims to maintain an accurate belief alongside its main goal, and it defines this type of problem as POMDP with \emph{informational rewards} (POMDP-IR). In this work, they propose expanding the action space with the estimation of the agent, and the cost function is modified to penalize/reward the incorrect/correct information. Additionally, in \cite{satsangi2018exploiting} it is shown that these two formulations can be converted to each other. 

In the information freshness literature, correlated sources are studied mainly in two settings. In the first setting, each sensor observes a set of processes, with possible overlaps, and the sensors update a single monitor \cite{chen2024improving, erbayat2024age}. Therefore, samples from different sensors may include asynchronous information about the same process, and received components have different freshness levels. In the second setting, there is a correlation between the processes observed by different sensors, thus getting an update from a sensor may contain partially fresh information about other processes \cite{he2018minimizing, tripathi2022optimizing, zancanaro2023modeling, hribar2017updating, liang2024optimizing, tong2022age, kumar2024age, he2019joint}, e.g., multiple cameras may surveil an overlapping area \cite{he2018minimizing}. Additionally, in \cite{fidler20242d}, the correlation is defined based on the spatial location of sources. Studies in \cite{he2018minimizing, tripathi2022optimizing, zancanaro2023modeling, hribar2017updating, liang2024optimizing, tong2022age, kumar2024age, he2019joint, fidler20242d} consider AoI of a correlated source as a weighted sum of the freshness of the corresponding sensors. In other words, when a source is sampled, AoI of other partially observed sources decreases proportionally with the correlation of the sampled source. In contrast, a metric called age of correlated information (AoCI) has been proposed in \cite{he2018minimizing}, and further studied in \cite{tong2022age, kumar2024age}, which considers a scenario where all sensors that include partial information about a process should be sampled in order to obtain fresh information about a process. Another related work \cite{tian2025real} considers a single source observed by multiple sensors, where each sensor observes the source process with different state-dependent accuracy. In this work, a time-invariant mismatch metric is minimized. We note that the setting of this paper involving monitoring partial states of a joint Markov process has not been studied before, to the best of our knowledge. 

In a pull-based scheme for a sensor network, the monitor can only be aware of the freshness of the sensor that has just been sampled, thus, POMDP formulation is a useful tool to model these types of problems. In \cite{shao2021partially, tahir2024collaborative, gong2020age, liu2023optimizing, liu2023optimizing2, chen2022uncertainty}, the monitor estimates the freshness of sensors that were not sampled at that slot based on its observations and obtains a policy accordingly. In \cite{shao2021partially}, \cite{tahir2024collaborative}, \cite{gong2020age}, and \cite{stamatakis2021autonomous}, sub-optimal policies are obtained via the index policy, the myopic policy, deep reinforcement learning, and the particle filter, respectively. The authors of \cite{liu2023optimizing} and \cite{liu2023optimizing2} convert the POMDP into a fully-observable discrete-space MDP problem, and obtain an optimum policy. Furthermore, \cite{zhao2025optimizing} and \cite{chiariotti2025goal} consider a push-based scheme where sensors estimate the freshness level of others to decide when to transmit. 

Reference \cite{he2024age} investigates a multi-agent optimization problem where agents should have \emph{fresh} and \emph{accurate} information about other agents' status to take better actions. Consequently, \cite{he2024age} utilizes the AoII metric and proposes to use a decentralized POMDP (Dec-POMDP) formulation which aims to minimize the AoII of each agent maintained for other agents' states. Similarly, in \cite{emami2024age}, uncrewed aerial vehicles (UAV) coordinate by estimating the states, e.g., speed, location, etc., of other UAVs, to collect fresh data from the ground sensors.

In early studies, AoII is investigated for symmetric Markov chains, and a single threshold policy is proposed to minimize the average AoII \cite{maatouk2020, chen2021minimizing, kriouile2022pull}. On the other hand, in our previous works \cite{cosandal2024modelingC, cosandal2024aoiiC, cosandal2024multi}, we have shown that if the source process is a general asymmetric Markov chain, a simple threshold policy would not be guaranteed to perform optimally. More specifically, in \cite{cosandal2024aoiiC, cosandal2024multi} it is shown that the optimum transmission policy should take into account all the estimation, source, and age values.  

The most commonly used remote estimation rule in the AoII literature is the so-called \emph{martingale estimator} \cite{akar2024query}, which estimates the process with the value encapsulated in the latest received status update from the information source \cite{maatouk2020, maatouk2022age, chen2021minimizing, kriouile2021minimizing, kam2020age}. However, this estimation rule is not applicable to our case, since we only get a partial observation with a status update. In a recent work \cite{cosandal2024joint}, a MAP rule is proposed for the estimation that allows the monitor to update its estimation with the most likely state, and it is shown that the MAP rule is superior to the martingale estimator for AoII minimization for Markov sources.

\section{System Model} \label{sec:SM}
We consider a time-slotted remote estimation system with $K$ sensors, where sensor-$k$, $k \in \mathcal{K}=\{1,2,\dots,K\}$, observes a discrete-time process $x_t(k) \in \mathcal{X}_k$, $|\mathcal{X}_k|=N_k$. Our main assumption is that these $K$ individual partial processes are not necessarily Markov, but the joint collection of these processes denoted by
\begin{align}
    \bm{x}_t=\{ x_t(1), x_t(2), \ldots, x_t(K) \}, 
\end{align}
is a discrete-time joint Markov process (or Markov chain). We denote the state-space of the joint process by $\mathcal{X}$. For convenience, we use an index process $\overline{x}_t\in\mathcal{N}=\{1,2,\dots, N\}$ that takes the index (or order) of the state of the original process $\bm{x}_t$ as its value, after a suitable enumeration of the joint states of the original process $\bm{x}_t$; see Fig.~\ref{fig:sys} for an example scenario. The index process $\overline{x}_t$ evolves according to a time-homogeneous transition matrix $\bm{P}=\{p_{ij}\}$, which is a-priori known, where $p_{ij}$ denotes the transition probability from index-state $i$ to index-state $j$. 

The remote monitor's goal is to track the joint process $\bm{x}_t$, equivalently its corresponding index process $\overline{x}_t$. At each time slot, the monitor decides whether to observe one of the sensors or stay idle. We denote the action taken by the monitor at time $t$ with $a_t=k$, $k\in\mathcal{K}\cup \{0\}$, where the action $0$ refers to staying idle, and action $k>0$ refers to sending a pull request to sensor-$k$ which is assumed to take place instantaneously. With the \emph{generate-at-will} (GAW) principle, if $a_t = k \neq 0$, then the sensor-$k$ transmits the value of the sampled process $x_t(k)$ via the forward channel encapsulated in an information packet, with the sampling and transmission cost denoted by $\mu_k > 0$. The forward channel is assumed to be an erasure channel with one-slot transmission time. In other words, the packet containing the value sampled from sensor-$k$ at time $t$, $x_t(k)$, is subject to a time-homogeneous transmission error with erasure probability $\rho_e$, i.e., the information packet is received at time $t+1$ by the monitor with successful transmission probability $\rho_s=1-\rho_e$. 

We denote the observation at the monitor at time $t$ with $o_t$ when action $a_{t-1}=k$, $k\in\mathcal{K}$, is taken at the previous time slot as,
\begin{align}
    o_t=\begin{cases}
        x_{t-1}(k), & \text{w.p.} \quad \rho_s, \\
        \emptyset,  & \text{w.p.} \quad \rho_e,
    \end{cases} \label{eq:obs}
\end{align}
where $o_t=x_{t-1}(k)$ denotes the received partial state value from sensor-$k$ and $\emptyset$ represents the channel error in the previous time slot. Additionally, the observation space for action $a_t=k$ indicates all possible observations from sensor-$k$, thus it is defined as $\mathcal{O}(k)=\mathcal{X}_k \cup \{\emptyset\}$. On the other hand, if the monitor stays idle, i.e., $a_t=0$, no pull request is sent to any sensor at time $t$. In this case, the observation space for $a_t=0$ is $\mathcal{O}(0)=\{\emptyset\}$. Thus, at the next time slot, the observation will be $o_t=\emptyset$ with probability (w.p.) one. Additionally, the sampling cost corresponding to staying idle is assumed to be zero, i.e., $\mu_0=0$. The described system model is illustrated in Fig.~\ref{fig:sys}. From all previous actions and received observations, the monitor can calculate the conditional probability distribution of the index process as a row vector of size $N$, 
\begin{align} 
    \bm{\pi}_t=\begin{bmatrix} \pi_t(1) & \pi_t(2) & \cdots & \pi_N(t) \end{bmatrix}, 
\end{align} 
where $\pi_t(i)$ corresponds to the conditional probability of the index process being in index-state $i$, 
\begin{align}
    \pi_t(i)=\mathbb{P}(\overline{x}_t=i|\mathcal{H}_t), \label{eq:pi}
\end{align}
and $\mathcal{H}_t$ is defined as the history of all observations and actions until time $t$ (excluding the action at time $t$), 
\begin{align}
    \mathcal{H}_t=\{o_1,a_1,\dots,a_{t-1},o_t\}. \label{eq:Hist}
\end{align}
The monitor estimates the index process using the MAP rule,
\begin{align}
    \hat{{x}}_t=\arg\max \bm\pi_t, \label{eq:map}
\end{align}
and the mismatch is quantified by AoII which progressively penalizes the incorrect estimation as the error stays. Particularly, the AoII process at time slot $t$, denoted by $\text{AoII}_t$, is a function of both the actual state $\overline{x}_t$ and its MAP estimation at the monitor $\hat{x}_t$, and $\text{AoII}_t$ evolves according to the following equation,
\begin{align}
    \text{AoII}_t=\begin{cases}
        \text{AoII}_{t-1}+1, & \hat{x}_t \neq \overline{x}_t, \\
        0, & \hat{x}_t = \overline{x}_t.
    \end{cases} \label{eq:MAoII}
\end{align}

We are interested in two types of costs in our setting. The first cost is MAoII, which is the time average of AoII due to the ergodicity of the corresponding AoII process,
\begin{align}
    \text{MAoII}=\lim_{T\to\infty} \frac{1}{T}\sum_{t=1}^T\text{AoII}_t.
\end{align}
The second cost is the average sampling cost (ASC), which can be expressed as a function of actions, 
\begin{align}
    \text{ASC} =\lim_{T\to\infty} \frac{1}{T}\sum_{t=1}^T \mu_{a_t}.
\end{align}

Finally, we define the optimization problem that aims to minimize the weighted sum of MAoII and $\text{ASC}$ as,
\begin{mini}
 	 	{\phi \in \Phi}{ \text{MAoII}^{\phi} + \lambda \cdot \text{ASC}^\phi, } 
 	{\label{Opt_eq}}
     {}
\end{mini}
where the coefficient $\lambda \geq 0$ represents the relative weight of average sampling cost to average age cost, $\phi$ is any monitor pull policy, and $\Phi$ is the set of all such policies. It is worthwhile to note that one can use the solution of this unconstrained problem for the solution of the MAoII minimization problem under average sampling cost constraints, with a proper selection of the coefficient $\lambda$ using an appropriate line search algorithm as in \cite{cosandal2024joint}.

\section{Problem Formulation} \label{sec:BMDP}
The monitor aims to find the optimum policy described in \eqref{Opt_eq} based on its previous actions and the observations it receives. It is important to note that the monitor will never have a perfect estimate of the joint process $\bm{x}_t$ due to communication delays, erasures, and the partial observability of the state, and consequently, it will not perfectly know the instantaneous value of AoII$_t$ for any value of $t$. Instead, the monitor keeps track of the joint distribution of index-state and AoII by using all observations it has received so far, which is termed as \emph{belief}. In this section, we utilize the belief to formulate the optimization problem as a belief-MDP. For this purpose, we first elaborate on the estimation update method introduced in the previous section. Then, we define the belief for this problem, and explain how it evolves with the observations. Lastly, the belief-MDP formulation is developed using the transition probabilities among beliefs under an action, and the expected cost of a belief-action pair.

\subsection{Estimation Update} \label{sec:EUpd}
In this subsection, we will describe how the estimation is updated as observations are made. For this purpose, in the first step, we assume that $\bm{\pi}_{t}$ is available to the monitor, and the new observation $o_{t+1}$ arrives. From \eqref{eq:obs}, if the packet is not lost during transmission, the new observation includes information about $\hat{x}_{t}$, thus $\bm\pi_{t}$ should be updated accordingly. Let us define the updated probability vector
\begin{align} 
    \hat{\bm{\pi}}_t=\begin{bmatrix} \hat{\pi}_t(1) & \hat{\pi}_t(2) & \cdots & \hat{\pi}_N(t) \end{bmatrix}, 
\end{align} 
where 
\begin{align}
    \hat{{\pi}}_{t}(i)=\mathbb{P}(\overline{x}_{t}=i|\mathcal{H}_{t+1}). \label{eq:hat_pi}
\end{align}
The following relation holds between ${\bm\pi}_{t}$ and $\hat{\bm\pi}_{t}$ for a given observation $o_{t+1}$,  
\begin{align}
\hat{\pi}_{t}(i)=
    \dfrac{ \mathbbm{1}_{o_{t+1},i} \pi_{t}(i)}{\sum_{j=1}^N \mathbbm{1}_{o_{t+1},j} \pi_{t}(j)}, \label{eq:hat_pi_t}
\end{align}
where $\mathbbm{1}_{o_{t+1},i}=\mathbb{P}(o_{t+1}|\overline{x}_{t}=i)$ is a binary function which indicates whether the observation $o_{t+1}$ is possible from $\overline{x}_t$. From the example process in Fig.~\ref{fig:sys}, $\mathbbm{1}_{a,i}$ takes the value $1$ when $i=1$ and $i=3$, because both states $(a,\alpha)$, $(a,\beta)$ are the only states possessing partial information `$a$' from the observation, and their indices are $1$ and $3$, respectively. Notice that an empty observation $o_t=\emptyset$ is possible for each state which makes $\hat{\pi}_{t}(i)={\pi}_{t}(i)$ for all values of $i$ when $o_t=\emptyset$.
In the second step, we obtain $\bm\pi_{t+1}$ from $\bm{\hat\pi}_{t}$ as follows,
\begin{align}
    \bm\pi_{t+1}=\bm{\hat\pi}_{t}\bm{P}. \label{eq:pi_t}
\end{align}
Once we obtain $\bm\pi_{t+1}$, the estimate $\hat{{x}}_t$ can be obtained from \eqref{eq:map}.

\subsection{Belief Evolution} \label{sec:BUpd}
In this subsection, we first define the belief $b_t$ as the following joint distribution
\begin{align}
    b_t(i,\delta)=\mathbb{P}(\overline{x}_t=i, \text{AoII}_t=\delta | \mathcal{H}_t), \label{eq:b}
\end{align}
and then describe how the belief evolves as observations are made. Additionally, in this paper, we truncate the AoII values by $\Delta$ for practical reasons.

To describe the evolution of a belief, we first define the 
\emph{updated belief} $\hat{b}_{t}$ as
\begin{align}
    \hat{b}_{t}(i,\delta)&=\mathbb{P}(\overline{x}_{t}=i, \text{AoII}_{t}=\delta|\mathcal{H}_{t+1}), \\
    &=\dfrac{ \mathbbm{1}_{o_{t+1},i} \, b_{t}(i,\delta)}{\sum_{d=0}^{\Delta}\sum_{j=1}^N \mathbbm{1}_{o_{t+1},j} \, b_{t}(j,d)}. \label{eq:hat_b_t} 
\end{align}
As \eqref{eq:MAoII} indicates, if a state transition is incurred to the estimated state, AoII resets to zero. Otherwise, the AoII process is incremented by one. In addition, we assume that the AoII value does not further increase upon reaching $\Delta$ that comes from truncation. As a result of this, the belief at time $t+1$ can be calculated using the updated belief, $\hat{b}_t$, and the source dynamics as follows,
\begin{align}
&b_{t+1}(i,\delta)=
\begin{cases}
        \max  \bm{\pi}_{t+1}, &  i=\hat{x}_{t+1}, \delta=0, \\
       \begin{aligned} \sum_{j=1}^{N}\hat{b}_{t}(j,\delta\!-\!1)p_{ji},\end{aligned} &  i\neq\hat{x}_{t+1}, 0 < \delta < \Delta,\\
        \begin{aligned}
            &\sum_{j=1}^{N}\hat{b}_{t}(j,\delta\!-\!1)p_{ji} \\
            &+\sum_{j=1}^{N}\hat{b}_{t}(j,\delta)p_{ji},
        \end{aligned}
         &  i\neq\hat{x}_{t+1},  \delta=\Delta,\\
        0, & \text{o.w.}
    \end{cases} \!\label{eq:b_t}
\end{align}
 Notice that when the estimation is correct at time $t+1$, i.e., $i=\hat{x}_{t+1}$, $\text{AoII}_{t+1}$ becomes zero regardless of its previous value, and its probability can be obtained from \eqref{eq:pi_t}. 

\begin{figure}[t]
    \centering
    \includegraphics[width=0.99\linewidth]{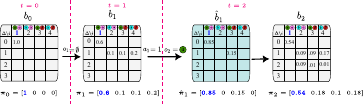}
    \caption{Evolution of the belief from an initial belief $b_0$ for the process in Fig.~\ref{fig:sys} with the transition matrix in \eqref{eq:Pexample}.}    \label{fig:example}
\end{figure}

\paragraph*{Example}
In this example, we consider again the process considered in Fig.~\ref{fig:sys}, with the transition matrix 
\begin{align}
    \bm{P}=\begin{bmatrix}
     0.6 & 0.1 & 0.1 & 0.2 \\
     0.2 & 0.6 & 0.1 & 0.1 \\
     0.1 & 0.2 & 0.6 & 0.1 \\
     0.1 & 0.1 & 0.2 & 0.6
\end{bmatrix}. \label{eq:Pexample}
\end{align}
We assume that at time $t=0$, the process is in state $(a,\alpha)$, i.e., the index process is in state 1, and this information is available to the monitor. Hence, $\bm{\pi}_0(1)=1, \bm{\pi}_0(i)=0$ for $i \neq 1$. Also notice that $b_0(1,0)=1, b_0(i,\delta)=0$ for $(i,\delta) \neq (1,0)$. We further assume that action 0 is taken at time $t=0$. Therefore, 
$\bm{\pi}_1=\begin{bmatrix}
    0.6 & 0.1 & 0.1 & 0.2
\end{bmatrix}$, and from \eqref{eq:map}, the estimation at the monitor at time $t=1$ is $\hat{x}_1=1$. 
Meanwhile, the AoII process will remain at zero with probability $0.6$, or it increases to $1$ with probability $0.4$, which makes $\mathbb{E}[\text{AoII}_1]=0.4$, and the joint state-AoII belief $b_1$ is obtained as in Fig.~\ref{fig:example}. Now, consider that the monitor decides to send a pull request to sensor-$1$ at time $t=1$, and the observation $o_2=a$ is successfully received at time $t=2$, which means that the unobserved state at $t=1$ was either $(a,\alpha)$ or $(a,\beta)$. Thus, the updated belief $\hat{b}_1$ is obtained by \eqref{eq:hat_b_t} as $\hat{b}_1(1,0)=0.85$, $\hat{b}_1(3,1)=0.15$, and $\hat{b}_1(i,\Delta)=0$, $(i,\Delta)\neq (1,0),(3,1)$. From that observation, the monitor can construct $b_2$ from \eqref{eq:b_t}, as illustrated in Fig.~\ref{fig:example}. A more detailed example with other possible observations is illustrated in Fig.~\ref{fig:table}.

It is well-known that belief $b_t$ contains all necessary information about history $\mathcal{H}_t$ regarding decision making, and thus, it is a sufficient statistic \cite{smallwood1973optimal}. Moreover, the new belief $b_{t+1}$ conditioned on the entire history of actions and observations until time $t+1$ can exactly be obtained from the previous belief $b_t$ along with the new observation $o_t$ using the equations  \eqref{eq:hat_b_t} and \eqref{eq:b_t} for which we say $b_{t+1}=f(b_t,o_{t+1})$, i.e., the current belief and the next observation (outcome of the current action) are mapped to the next belief,  which is represented by the mapping (i.e., the function) $f$. 
In the previous example illustrated in Fig.~\ref{fig:example}, $b_2$ is obtained from $b_1$ and the observation $o_2=a$, thus $b_2=f(b_1,a)$. 

\subsection{Belief-MDP Formulation}
For the belief-MDP formulation, we first define $T(b_{t+1},a_t,b_t)$ as the transition probability from $b_t$ to $b_{t+1}$ for a given action $a_t$. Although the belief space includes continuous-valued elements, the number of possible next beliefs is limited by the number of possible observations. In other words, for any belief $b_t$ and action $a_t$, there are $|\mathcal{O}(a_t)|$ possible next beliefs $b_{t+1}$. Additionally, transition probability between $b_t$ and $b_{t+1}=f(b_{t},o_{t+1})$ for an action $a_t$ is equivalent to the probability of the observation $o_{t+1}$ for given $b_t$ and $a_t$, since $f(b_{t},o_{t+1})$ is a deterministic mapping. This can be expressed as
\begin{align}
    T(f(b_{t},o_{t+1}),b_t,a_t)=&\mathbb{P}(b_{t+1}=f(b_{t+1},o_{t+1})|b_t,a_t)\\=&\mathbb{P}(o_{t+1}|b_t,a_t). \label{eq:Tp}
\end{align}

The observation $o_{t+1}=\emptyset$ is included in the observation space of any action, including $a_{t}=0$. If the action is $a_{t}\neq0$, the observation $o_{t+1}=\emptyset$ denotes the erasure on the channel, and it occurs with the probability of $\rho_e$. Additionally, if the monitor stays idle with action $a_{t}=0$, it always observes $o_{t+1}=\emptyset$. Thus, we can express these probabilities as
\begin{align}
    \mathbb{P}(o_{t+1}=\emptyset|a_{t},b_{t})&= \begin{cases}
    \rho_e, & a_t\neq 0, \\
    1, & a_t=0.
    \end{cases} \label{eq:tprob_e}
\end{align}

Furthermore, the observation $o_{t+1}\neq\emptyset$ indicates the following: i) monitor sends the pull request to one of the sensors at time $t$, i.e. $a_t=k\neq0$, ii) by \eqref{eq:obs} the observation $o_{t+1}$ is included the observation space of the action, i.e., $o_{t+1}\in\mathcal{O}(k)$,  iii) transmission succeeds with probability $\rho_s$, iv) the observation at time $t+1$ is the $k$th component of the process at time $t$, i.e., $o_{t+1}=x_t(k)$.  We can collect all these conditions with the following equation:
\begin{align}
    &\mathbb{P}(o_{t+1}\neq\emptyset|a_{t},b_{t})\nonumber\\
    &=\rho_s\sum_{i=1}^N \mathbbm{1}_{o_{t+1},i} \mathbb{P}(\overline{x}_{t}=i|\mathcal{H}_{t+1}), \quad a_t\neq0, \ o_{t+1}\in\mathcal{O}(a_t),
    \label{eq:tprob_s}
\end{align}
Notice that statements (i) and (ii) are denoted with $a_t\neq0$, $o_{t+1}\in\mathcal{O}(a_t)$, respectively. The successful transmission statement in (iii) adds the multiplication term $\rho_s$ on the probability, and it is denoted with $o_{t+1}\neq\emptyset$. The last statement in (iv) is expressed with $\sum_{i=1}^N \mathbbm{1}_{o_{t+1},i} \mathbb{P}(\overline{x}_{t}=i|\mathcal{H}_{t+1})$ which is equivalent to $\sum_{i=1}^N \mathbbm{1}_{o_{t+1},i} \pi_t(i)$, or $\sum_{i=1}^N \mathbbm{1}_{o_{t+1},i}\sum_{\delta=0}^{\Delta} b_t(i,\delta)$. Finally, we can combine \eqref{eq:Tp}, \eqref{eq:tprob_e} and \eqref{eq:tprob_s} to obtain the function $T$ as
\begin{align}
    &T(f(b_{t},o_{t+1}),b_t,a_t)\nonumber\\&=\begin{cases}
        \rho_s\sum_{i=1}^N \sum_{\delta=0}^{\Delta}\mathbbm{1}_{o_{t+1},i} b_t(i,\delta), & o_{t+1}\in\mathcal{O}(a_t), o_{t+1}\neq\emptyset,\\
        \rho_e, & o_{t+1}=\emptyset, \ a_t\neq0\\
        1, & a_t=0, \\
        0, & \text{o.w.}
    \end{cases} \label{eq:tprob}
\end{align}
Notice that the conditions $o_{t+1}\in\mathcal{O}(a_t)$, and $o_{t+1}\neq\emptyset$ inherently indicate that the action is non-zero, thus the condition $a_t\neq0$ is simplified from the first case of the equation.

\begin{figure}[t]
    \centering
    \includegraphics[width=0.95\linewidth]{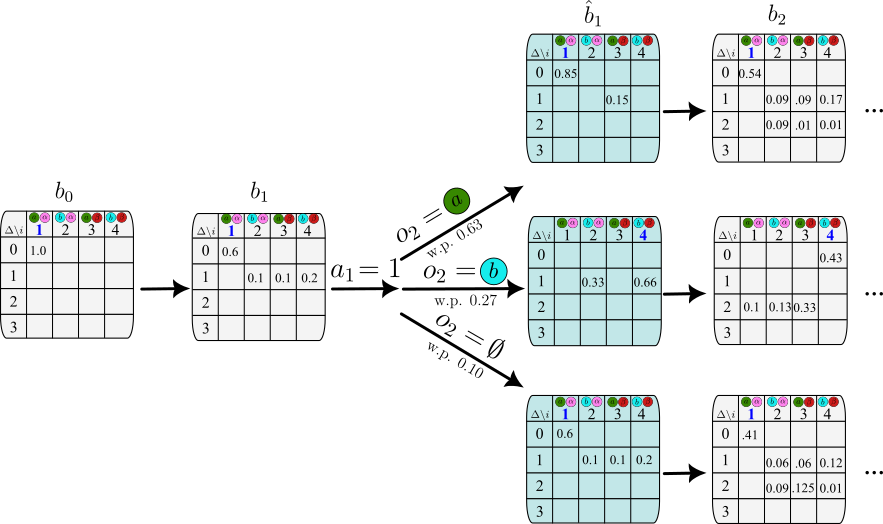}
    \caption{Illustration of all possible outcomes $\hat{b}_1$, $b_2$ and the probability of the corresponding observations when the action $a_1=1$ is chosen for $\rho_e=(1-\rho_s)=0.1$.}
    \label{fig:table}
    \vspace*{-0.4cm}
\end{figure}

In the example in Fig.~\ref{fig:table}, the action $a_{1}=1$ is taken at time $t=1$. This action has the observation space $\mathcal{O}(1)=\{\emptyset,a,b\}$. The observations `$a$', `$b$' include partial information for states $\overline{x}_1=1,3$, and $\overline{x}_1=2,4$, respectively, thus $\mathbbm{1}_{a,i}=1$ for $i=1,3$, and $\mathbbm{1}_{b,i}=1$ for $i=2,4$. From the given belief $b_1$, the transmission probabilities to all possible $b_2$ values are calculated by \eqref{eq:tprob} for the corresponding observation, and denoted under the corresponding arrows.

The expected cost of the belief state $b_t$ and action $a_t$ is defined as the sampling cost of the action plus the expected AoII of the next reached belief. First, we can calculate the expected AoII of a belief $b_t$ by its definition in \eqref{eq:b}, and denote it with a function of $b_t$ as
\begin{align}
   \Gamma(b_t)&= \mathbb{E}[\mbox{AoII}_t]=\sum_{\delta=0}^{\Delta} \delta \sum_{i=1}^N\mathbb{P}(\overline{x}_t=i,\mbox{AoII}_t=\delta|\mathcal{H}_t),\\
    &=\sum_{\delta=0}^{\Delta}\delta\sum_{i=1}^N b_t(i,\delta).
\end{align}
Then, the expected cost is defined as
\begin{align}
     &c(b_t,a_t)=\nonumber\\&\sum_{o_{t+1}\in\mathcal{O}(a_t)}T(f(b_{t},o_{t+1}),b_t,a_t)\Gamma(f(b_t,o_{t+1}))+\lambda \mu_{a_t}. \label{eq:cost}
\end{align}

Finally, we summarize the belief-MDP by using formulations in \cite{dagan2024resolving, lim2023optimality} as a tuple $\left(\mathcal{B},\mathcal{A},T,c,\gamma\right)$:
\begin{itemize}
    \item The belief space is $\mathcal{B}=[0,1]^{ N(\Delta+1)}$, and the belief of the problem, $b_t \in \mathcal{B}$ is defined in \eqref{eq:b}.
    \item  The action space of the problem is defined as $\mathcal{A}=\mathcal{K}\cup\{0\mbox\}$.
    \item  After the reception of an observation, the state distribution is updated using  \eqref{eq:hat_pi_t} and \eqref{eq:pi_t}, and the estimation is updated by \eqref{eq:map}.
    \item Each observation $o_{t+1}$ forces the belief $b_t$ according to $b_{t+1}=f(b_t,o_{t+1})$ with probability $T(f(b_t,o_{t+1}),a_t,b_t)$. The derivation of $f(b_t,o_{t+1})$ is detailed in the steps \eqref{eq:hat_b_t} and \eqref{eq:b_t}, and the transition probability is derived in \eqref{eq:tprob}.
    \item  The cost of a belief-action pair $(b_t,a_t)$  is defined through a cost function $c(b_t,a_t)$ in \eqref{eq:cost}.
    \item  It is a common practice to choose the discount factor as $\gamma=1$ \cite{boccia2014stability,lin2023reinforcement, bertsekas2024model, sehr2018performance, ulfsjoo2022integrating, esfahani2021reinforcement}. Hence, this term is omitted in the rest of the paper.
\end{itemize}

\section{Model Predictive Control} \label{sec:MPC}
\begin{figure}[t]
    \centering
    \includegraphics[width=0.95\linewidth]{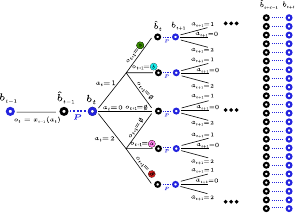}
    \caption{An $\ell$-step look-ahead table that starts a belief state and explores all possible belief states on $\ell$ steps.}
    \label{fig:rollout}
\end{figure}

The belief-MDP formulation defined in the previous section has a continuous state-space and a discrete action space. Continuous state-space disallows us to find an optimum policy for the optimization problem in \eqref{Opt_eq} via dynamic programming. On the other hand, the discrete action space indicates that for a single belief, there are finite number of next beliefs. Therefore, we utilize the MPC where the monitor searches the sequence of actions $\bm{u}^*_t=[u_t(1),\dots,u_{t}(\ell-1)]$, which minimizes the expected cost by starting with a belief $b_t$, and evaluating all possible reachable beliefs using an $\ell$-step look-ahead table. Then, the monitor applies the first action from the action sequence, i.e., $a_t^*=u^*_t(1)$. An example of the look-ahead table is given in Fig.~\ref{fig:rollout}. The action sequence $\bm{u}^*_t$ is expressed as
\begin{align}
    \underset{\substack{a_{t+m}\in \mathcal{A}, \\ m\in\{0,\dots,\ell-1\}} }{\arg\min} \mathbb{E}\Bigg[\sum_{m=0}^{\ell-1} c(b_{t+m},a_{t+m})+V(b_{t+\ell-1})\nonumber \\ \bigg|b_t,a_{t},\dots,a_{t+\ell-1}\Bigg], \label{eq:smpc}
\end{align}
where $V(b_{t+\ell-1})$ is the terminal cost of the belief reached after $\ell$ actions. We propose two MPC approaches based on two terminal cost assumptions.

\subsection{MPC-WTC}
In MPC-WTC, terminal costs of beliefs are fixed to $0$. Therefore, the monitor chooses the action based on the expected cost from the first $\ell$ steps. This expected value is calculated recursively by defining $C_{m,\ell}(b_k,a_k)$, the cost in next $\ell-m$ steps for the belief $b_k$ and the action $a_k$, as 
\begin{align}
    &C^{\text{WTC}}_{m,\ell}(b_k,a_k)=c(b_k,a_k)\nonumber\\&+\sum_{o_{k+1}\in\mathcal{O}(a_k)} T(f(b_k,o_{k+1}),a_k,b_k)  C^{\text{WTC}}_{m+1,\ell}(f(b_k,o_{k+1})), \label{eq:R_k}
\end{align}
where $C^{\text{WTC}}_{m,\ell}(b_k)$ is the value of the belief by considering the actions that minimize cost for the remaining steps are applied, equivalently 
\begin{align}
    C^{\text{WTC}}_{m,\ell}(b_k) = \underset{a_k\in\mathcal{A}}{\min} \ C^{\text{WTC}}_{m,\ell}(b_k,a_k), \quad m<\ell-1. \label{eq:R_b}
\end{align}
The value of $C^{\text{WTC}}_{\ell-1,\ell}(b_k)=V(b_{k})=0$ stands for the terminal cost, which is fixed to $0$. In each time slot $t$, the monitor chooses the action as $a^*_t=\arg\min{C^{\text{WTC}}_{0,\ell}(b_t,a)}$, and the value of $C^{\text{WTC}}_{0,\ell}(b_t,a)$ is obtained recursively by Algorithm~\ref{alg:rec}.

\begin{algorithm}[h]
    \caption{$C_{m,\ell}(b_k)$: The recursive algorithm to calculate the expected cost in the next $\ell-m$ steps.}\label{alg:rec}
    \begin{algorithmic}
    \For{$a_k \in \mathcal{A}$}          
    \For{$o_{k+1} \in \mathcal{O}(a_k)$}
    \If{$m=\ell-1$}
    \State $\sigma_{b_k,a_k,o_{k+1}}=V(b_k)$
    \Else
    \State $\sigma_{b_k,a_k,o_{k+1}}=C_{m+1,\ell}(f(b_k,o_{k+1}))$
    \EndIf
    \EndFor
    \EndFor
    \State \textbf{Return:} 
    $\min_{a_k\in\mathcal{A}} \ c(f(b_k,o_{k+1}),a_k)+$ \\ $\qquad \qquad \quad \sum_{o_{k+1}\in\mathcal{O}(a_k)}  T(f(b_k,o_{k+1}),a_k,b_k)\sigma_{b_k,a_k,o_{k+1}}$
    \end{algorithmic}
\end{algorithm}

\begin{figure}[!t]
    \centering
    \includegraphics[width=0.65\linewidth]{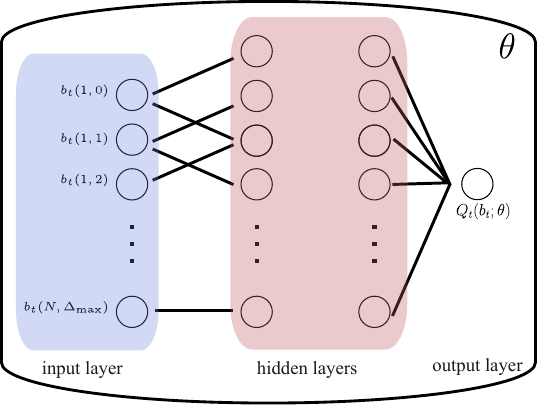}
    \caption{Network architecture used for RL-MPC with parameter $\theta$.}
    \label{fig:RLj}
\end{figure}

\subsection{RL-MPC}
For RL-MPC, we adapt the iterative method in \cite{lin2023reinforcement}, which in each iteration $d$, aims to approximate the minimum expected cost of the first $d\ell$ steps with a neural network as,
\begin{align}
   & Q(b_t;\theta_d)\approx \nonumber \\
   & \underset{\substack{a_{t+m}\in \mathcal{A}, \\ m\in\{1,\dots,\ell-1\}} }{\min} \mathbb{E}\left[\sum_{m=1}^{d\ell} c(b_{t+m},a_{t+m})\bigg|b_t,a_{t},\dots,a_{t+d\ell-1}\right],
\end{align}
where $\theta_d$, and $Q(b_t;\theta_d)$ denote the network parameters and the output of the network at iteration $d$, respectively. The neural network design is illustrated with Fig.~\ref{fig:RLj}, and the trained network in each iteration is utilized to calculate the terminal costs in the next iteration.

We implement RL-MPC as follows. We first initiate the network parameters to satisfy $Q(b_t;\theta_d)=0$, $b_t\in\mathcal{B}$ for $d=0,1$. Then, in each iteration, we train the network with the loss function 
\begin{align}
    J(\theta_d)=Q(b_t;\theta_d)-C^{RL}_{0,\ell,d}(b_t), \quad t\in \{0,1,\dots,T_{\max}\},
\end{align}
where $T_{\max}$ is the training horizon, and $C^{\text{RL}}_{m,\ell,d}(b_t,a_t)$ is defined similar to MPC-WTC as the cost in the next $\ell-m$ steps for the belief $b_t$ and the action $a_t$ in iteration $d$, and it has the recursive relation
\begin{align}
    &C^{\text{RL}}_{m,\ell,d}(b_k,a_k)=c(b_k,a_k)\nonumber\\&+\sum_{o_{k+1}\in\mathcal{O}(a_k)} T(f(b_k,o_{k+1}),a_k,b_k)  C^{\text{RL}}_{m+1,\ell,d}(f(b_k,o_{k+1})), \label{eq:R_k_RL}
\end{align}
Different from MPC-WTC, the network from the previous iteration is used to calculate terminal costs as $C^{\text{RL}}_{\ell-1,\ell}(b_k)=V(b_k)=Q(b_k;\theta_{d-1})$. Thus, $C^{\text{RL}}_{m,\ell,d}(b)$ can be expressed for RL-MPC as 
\begin{align}
    C^{\text{RL}}_{m,\ell,d}(b_k) = \begin{cases}
        \underset{a_k\in\mathcal{A}}{\min} \ C^{\text{RL}}_{m,\ell,d}(b_k,a_k), & m<\ell, \\
        Q(b_k;\theta_{d-1}), & m=\ell.
    \end{cases}  \label{eq:R_b_RL}
\end{align}
Next, the action $a^*_t=\arg\min{C^{\text{RL}}_{0,\ell,d}(b_t,a_t)}$ is applied, and the value of $C^{\text{RL}}_{0,\ell,d}(b_t,a_t)$ is obtained by Algorithm~\ref{alg:rec} by substituting $C_{m,\ell}(b_k)=C^{\text{RL}}_{0,\ell,d}(b_t)$, and $V(b_{k})=Q(b_k;\theta_{d-1})$ at iteration $d$. 

Notice that, in the first iteration, the algorithm works identically to MPC-WTC with the same look-ahead step size by learning $Q(b_t;\theta_1)\approx C^{\text{WTC}}_{0,\ell}(b_t)$, and in the remaining steps, the approximation horizon is extended. For instance, for $d=2$ and $\ell=1$, the network output approximates 
\begin{align}
    &Q(b_t;\theta_2)\approx\min_{a_t\in\mathcal{A}}\bigg[ C^{\text{WTC}}_{0,1}(b_t,a_t) \nonumber\\ & + \sum_{o_{t+1}\in\mathcal{O}(a_t)} T(f(b_t,o_{t+1}),a_t,b_t) Q(f(b_t,o_{t+1});\theta_1)\bigg]\\
    &\approx\min_{a_t\in\mathcal{A}}\bigg[ C^{\text{WTC}}_{0,1}(b_t,a_t) \nonumber\\ &
    + \sum_{o_{t+1}\in\mathcal{O}(a_t)} T(f(b_t,o_{t+1}),a_t,b_t) C_{0,\ell}^{\text{WTC}}\left(f(b_t,o_{t+1})\right)\bigg] \\ &\approx  C_{0,2}^{\text{WTC}}\left(b_t,a_{t}\right)
    .
\end{align}

\section{Numerical Results} \label{sec:NR}

\begin{table}
    \centering
    \caption{Parameters used in the simulation experiments.}
    \begin{tabular}[t]{|c|c|c|}
    \hline
    $d_{\max}$  & 4\\ \hline
    learning rate& $10^{-3}$\\ \hline
    numbers of hidden layers& 2\\ \hline
    nodes at hidden layers& $60$\\ \hline
    $\mu_k$ & $1$, for $k\in\mathcal{K}$ \\ \hline
    $\Delta$& 15\\ \hline
    $T_{\max}$ & $10^6$ \\ \hline
    \end{tabular}
    \label{tab:sim_tab}
\end{table}

We use the following open-loop policies as benchmark policies in order to compare their performance to the proposed MPC methods. For all benchmark policies, the sampling frequency is controlled with a single parameter $\alpha$, and the optimum $\alpha^*$ value is used for each benchmark policy, which minimizes the optimization problem in \eqref{Opt_eq} for a given value of $\lambda$. The method we use for finding the optimum $\alpha^*$ value for all benchmark policies is the use of simulations and brute force search. 

\paragraph{Benchmark I} 
In this benchmark policy, pull requests are generated with probability $\alpha$, and a generated request is directed towards each of the sensors in a uniformly likely manner. The action $a_t$ for Benchmark I taken at time $t$, can then be mathematically expressed as,
\begin{align}
    a_t^{\text{rand}} = 
    \begin{cases}
    n,  & \text{w.p. } \frac{\alpha}{N}, \   n \in \{1, \ldots,N\},\\
    0, & \text{w.p. } 1-\alpha.
    \end{cases}
\end{align}

\paragraph{Benchmark II} In this benchmark policy, pull requests are generated almost uniformly with a period ${1}/{\alpha}$. In other words, the $m$th pull request is generated at $\text{round}({m}/{\alpha})$th time slot, where the rounding operation, denoted by $\text{round}(\cdot)$, maps its argument to the nearest integer. In addition, the monitor chooses one of the sensors in a round-robin (i.e., circular) fashion. The action $a_t$ for this policy can be expressed as,
\begin{align}
    a_{t}^{\text{rr}} = 
    \begin{cases}
        \text{mod}(a_{\eta(t)}^{\text{rr}}, K) + 1, & \text{if } t \in \left\{ \text{round}\left(\frac{m}{\alpha}\right):\, m \in \mathbb{N} \right\}, \\
        0, & \text{o.w.}
    \end{cases}
\end{align}
Here, $\eta(t)$ is the time of the last transmission before $t$.

\paragraph{Benchmark III} This benchmark policy modifies the previous benchmark policy by incorporating \emph{erasure-awareness}, i.e., it repeats the previous action if the message is erased during transmission. The action for this case can be expressed mathematically as, 
\begin{align}
    a_{t}^{\text{rr,ea}} = 
    \begin{cases}
        \text{mod}(a_{\xi(t)}^{\text{rr,ea}}, K) + 1, & \text{if } t \in \left\{ \text{round}\left(\frac{m}{\alpha}\right):\, m \in \mathbb{N} \right\}, \\
        0, & \text{o.w.}
    \end{cases}
\end{align}
Here, $\xi(t)$ is the time  of the last \emph{successful} transmission before $t$.

In all simulations, we assume that the initial state is known by the monitor; hence, the initial AoII value is $0$. For all policies, AoII values are truncated to $\Delta=15$, and the MAoII values are obtained over at least $10^6$ state transitions starting from the initial index-state $\overline{x}_1=1$. Unless otherwise stated, we assume a homogeneous sampling cost for all sensors which is fixed to $1$. For RL-MPC, a neural network with two hidden layers is used with the learning rate of $10^{-3}$. We train the network for at least $d_{\max}=4$ iterations before applying it. These parameters are summarized in Table~\ref{tab:sim_tab}. Our numerical results pertaining to two different application scenarios are presented in the two subsections.

\subsection{Scenario I: Random Walk on a 2D Grid}

\begin{figure}
    \centering
    \includegraphics[width=0.65\linewidth]{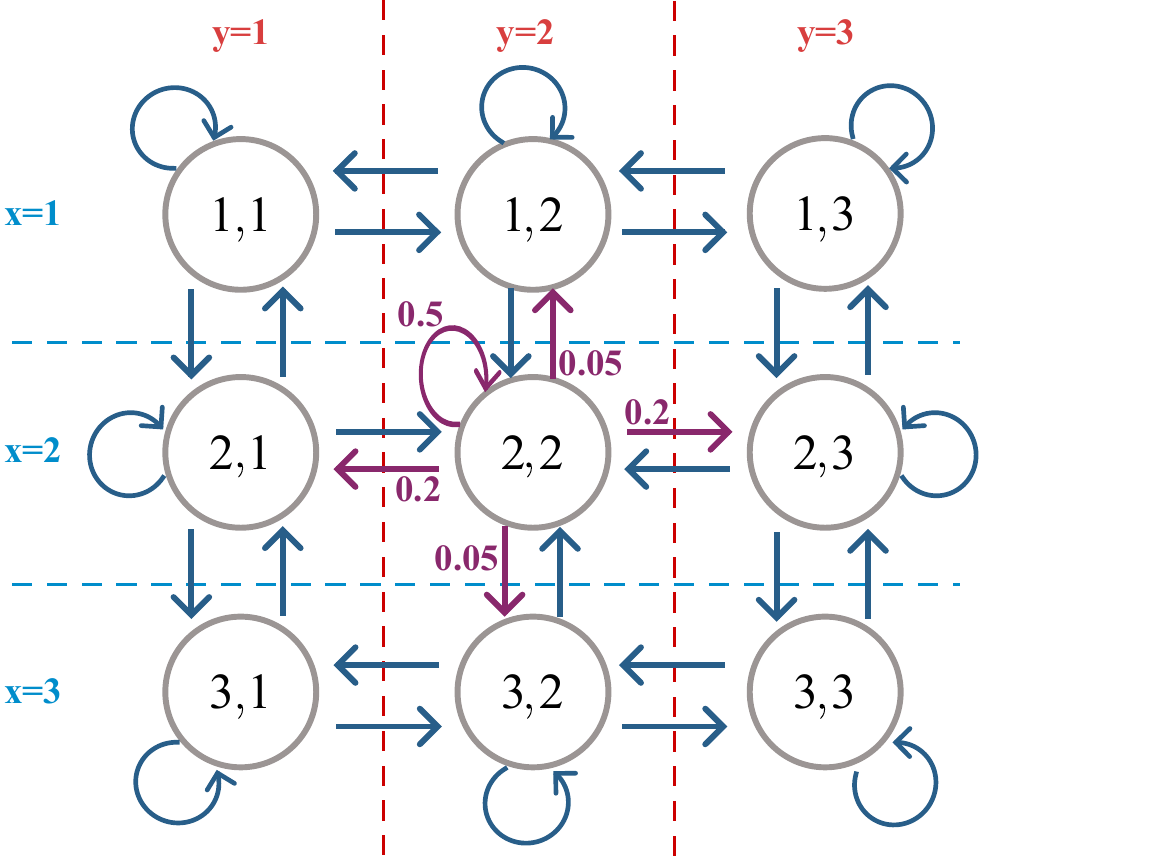}
    \caption{State diagram for a $3\times3$ grid. Only the transition probabilities from state $(2,2)$ are given.}
    \label{fig:grid_st}
\end{figure}

We consider an $L_x \times L_y$ grid space over which an object performs a random walk. We consider two sensors, sensor-$1$ and sensor-$2$, that are only able to observe the object's $x$ and $y$ coordinates, respectively. For the random walk dynamics, we choose the probability of staying in the same state as $0.5$, moving horizontally as $0.4$ (with equal probability for right and left), and moving vertically as $0.1$ (with equal probability for down and up). For the states at the corners and boundaries, these probabilities are normalized. An example of the state diagram for $N=3$ is illustrated in Fig.~\ref{fig:grid_st} with the $x$ and $y$ coordinates of all states.

\begin{figure}
    \begin{center}
    \subfigure[$\lambda=0$]{\includegraphics[width=0.8\columnwidth]{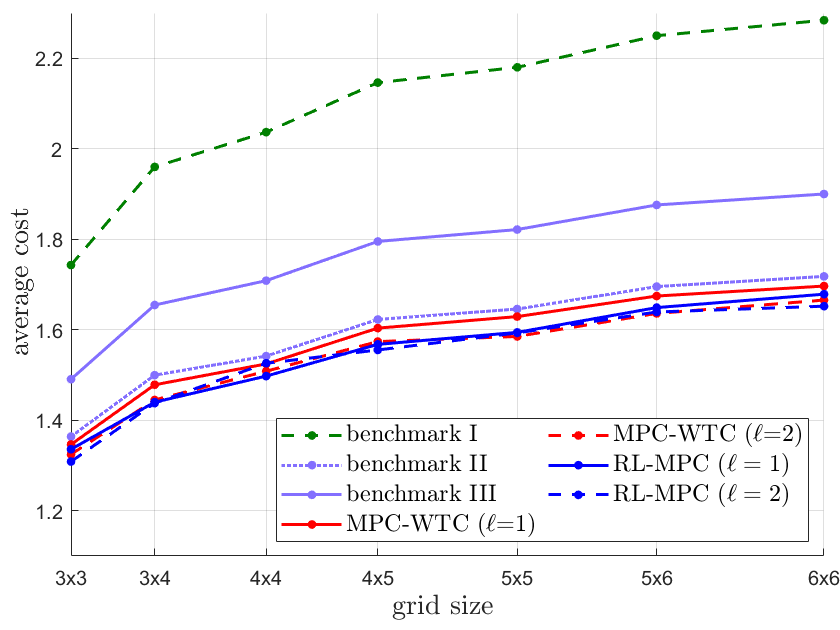}}
    \subfigure[$\lambda=0.5$]{\includegraphics[width=0.8\columnwidth]{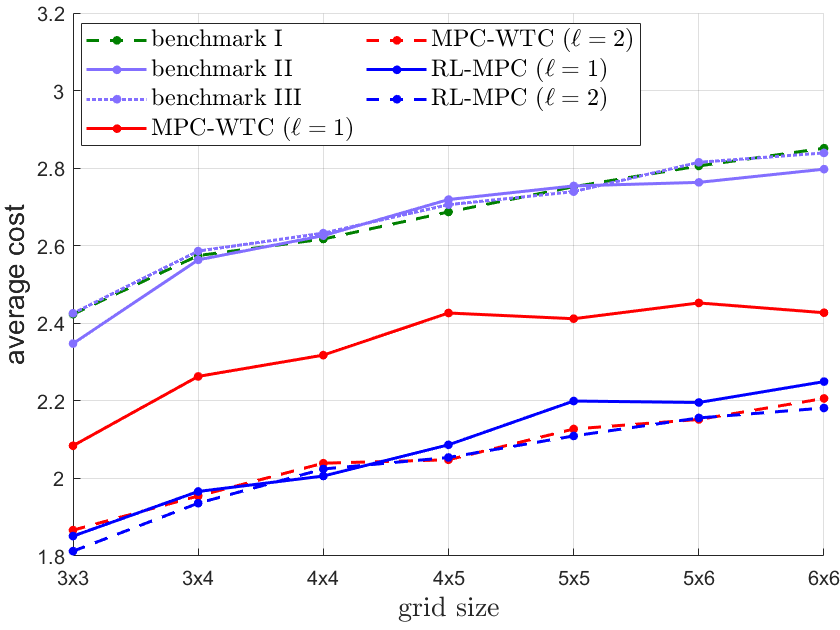}}  
    \end{center}      \caption{Comparison of methods for different grid sizes and $\lambda$ values when $\rho_s=0.8$.}
    \label{fig:grid}
    \vspace*{-0.4cm}
\end{figure}

In Fig.~\ref{fig:grid}, we compare the performance of various pull policies in terms of the average cost 
$\text{MAoII }+\lambda \cdot \text{ASC}$ for different grid sizes $L_x \times L_y$ while fixing $\rho_s=0.8$. We first observe that round-robin policies (benchmarks II and III) provide performance gains in comparison to random sampling (benchmark I), and this performance gain is larger with the erasure-aware scheme (benchmark III) for $\lambda=0$. However, as $\lambda$ increases, the gain attained with erasure awareness appears to diminish. Furthermore, MPC methods outperform all benchmark methods in this scenario. We observe that MPC-WTC performs substantially better with increasing the look-ahead step size from $\ell=1$ to $\ell=2$. On the other hand, RL-MPC attains the same performance with the choice $\ell=1$.

\begin{figure}
    \centering
    \includegraphics[width=0.8\linewidth]{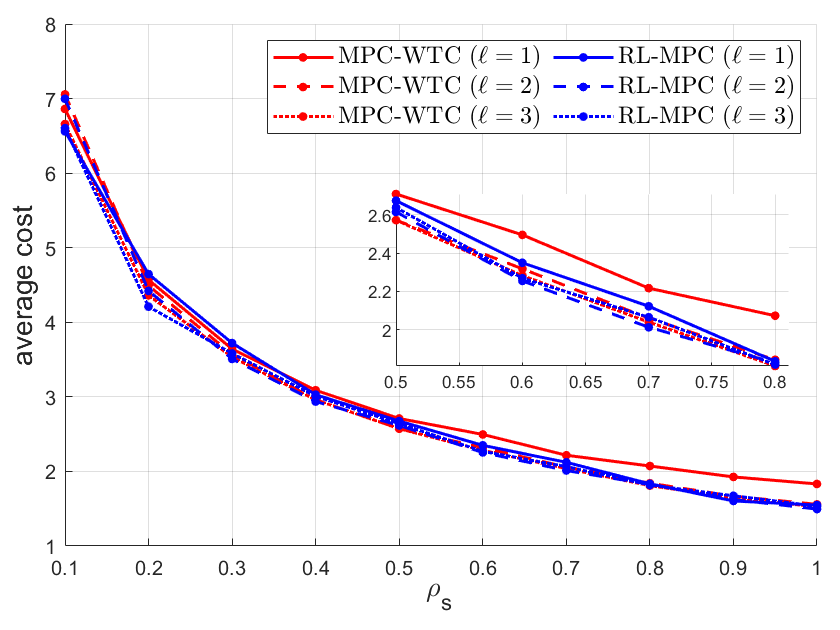}
    \caption{Comparison of MPC policies with different look-ahead step size and varying $\rho_s$ for $3\times3$ grid size.}
    \label{fig:grid3}
\end{figure}

\begin{figure}
    \centering
    \includegraphics[width=0.75\linewidth]{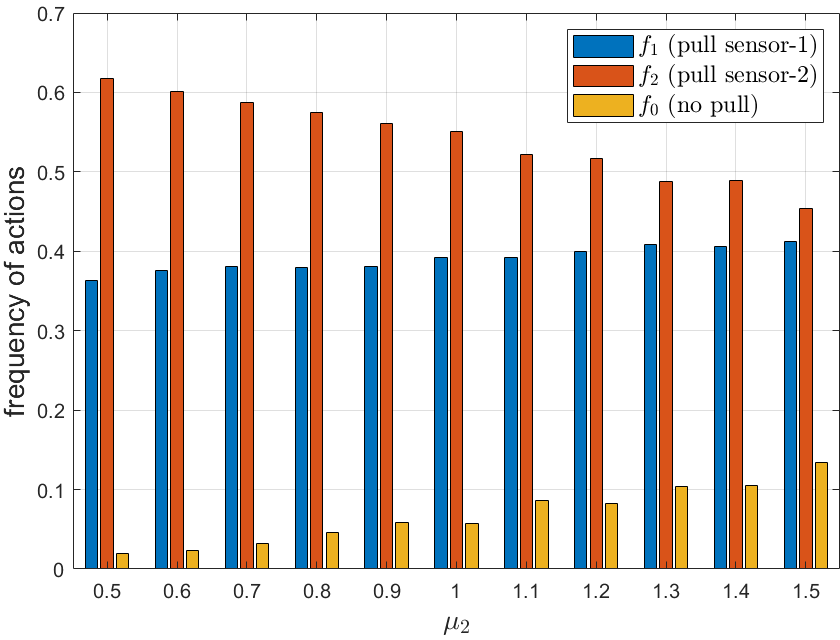}
    \caption{Frequencies of the actions for applying RL-MPC $(\ell=2)$ to $3\times3$ grid with varying sampling cost for sensor-$2$ ($\mu_2$) and fixed sampling cost for sensor-$1$ ($\mu_1=1$).}
    \label{fig:freq}
\end{figure}

In Fig.~\ref{fig:grid3}, we fix the grid to $L_x=L_y=3$ and compare the average cost of the MPC policies for different look-ahead step sizes as a function of $\rho_s$. We observe that increasing the look-ahead step size further from $\ell=1$ for RL-MPC, and from $\ell=2$ for MPC-WTC, does not substantially improve their performance in terms of the average cost.
In Fig.~\ref{fig:freq}, when $L_x=L_y=3$, we also fix the sampling cost for sensor-$1$ to $\mu_1=1$, and illustrate the frequency of actions in RL-MPC ($\ell=2$) when the sampling cost parameter of the second sensor is varied from $\mu_2=0.5$ to $\mu_2=1.5$. We denote the frequency of an action $u\in\mathcal{A}$ with $f_u$, which measures how often an action is taken for a given policy, and it is written as,
\begin{align}
    f_u=\frac{1}{T}\sum_{t=1}^T \mathbbm{1}_{u,a_t},
    \label{eq:freq}
\end{align}
where $\mathbbm{1}_{u,a_t}$ is the indicator function which is $1$ if $a_t$ is the same with action $u$, and $T$ is chosen as $T=10^6$ for this experiment. We observe that even when the sampling costs of both sensors are equal, i.e., $\mu_1=\mu_2=1$, the monitor tends to sample sensor-$2$ more frequently to track the location, as it is more likely that the object makes a horizontal move due to the dynamics of the joint process.  On the other hand, as $\mu_2$ increases, the frequencies at which sensor-$1$ is pulled, and no pull requests are made, also increase.

\subsection{Scenario II: Correlated Temperature/Fire/Freeze Events}

\begin{figure}
    \centering
    \includegraphics[width=0.65\linewidth]{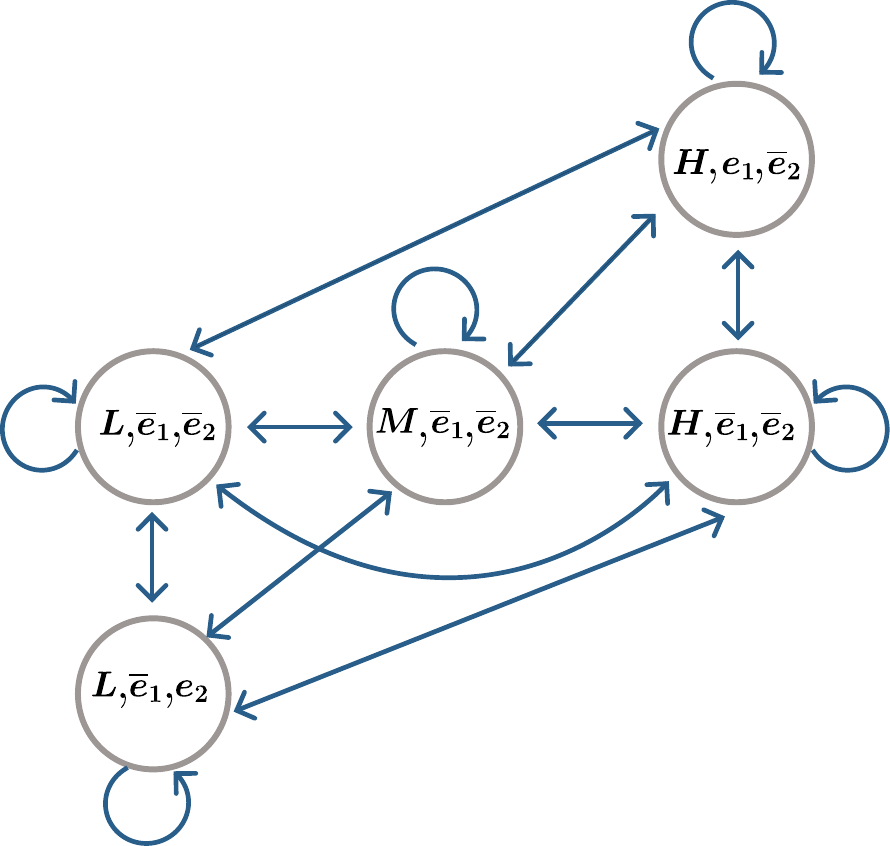}
    \caption{State diagram for scenario II. Bidirectional arrows and self-loops indicate non-zero transition probabilities between two states in either direction, and non-zero self-transition probabilities, respectively.}
    \label{fig:2x3x3}
\end{figure}

In this scenario, we consider a system model with three sensors. The first sensor observes the temperature of the environment. The temperature process has states `$H$', `$M$', and `$L$' corresponding to high, moderate, and low temperatures, respectively. The remaining two sensors observe events with state-space $\mathcal{X}_2=\{e_1,\tilde{e}_1\}$, and $\mathcal{X}_3=\{e_2,\tilde{e}_2\}$, and the event $e_1$ (resp. $e_2$) occurs only when the underlying process is in the state $x_t(1)=H$ (resp. $x_t(1)=L$). These temperature-dependent events can be considered as ``fire'' and ``freeze'', respectively. Therefore, the number of reachable joint states is reduced to $5$ as illustrated in Fig.~\ref{fig:2x3x3}, and these states are sorted as $\mathcal{X}=$ $\{(H,e_1,\tilde{e}_2)$, $(H,\tilde{e}_1,\tilde{e}_2)$, $(M,\tilde{e}_1,\tilde{e}_2)$, $(L,\tilde{e}_1, {e}_2)$, $(L,\tilde{e}_1,\tilde{e}_2)\}$. Note that obtaining observations $o_{t}=e_1$, $o_{t}=M$, and $o_{t}=e_2$ indicates that the index process is $\overline{x}_{t-1}=1$, $\overline{x}_{t-1}=3$, and $\overline{x}_{t-1}=5$, respectively. The transition probability matrix for the index process, denoted by $\bm{P}$, is chosen as,
\begin{align}
    \bm{P}=\begin{bmatrix}
    0.1 & 0.7 & 0.1 & 0.1 & 0 \\
    0.4 & 0.4 & 0.1 & 0.05 & 0.05 \\
    0.05 & 0.05 & 0.8 & 0.05 & 0.05 \\
    0.05 & 0.05 & 0.1 & 0.1 & 0.7 \\
    0.1 & 0 & 0.1 & 0.4 & 0.4 
    \end{bmatrix}.
\end{align}

\begin{figure}[t]
    \begin{center}
    \subfigure[$\lambda=0$]{\includegraphics[width=0.8\columnwidth]{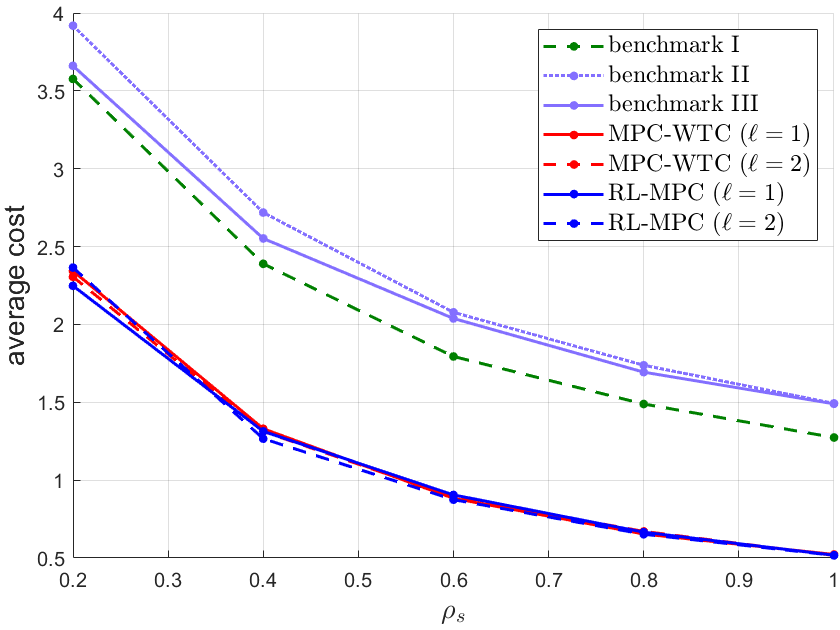}}
    \subfigure[$\lambda=1$]{\includegraphics[width=0.8\columnwidth]{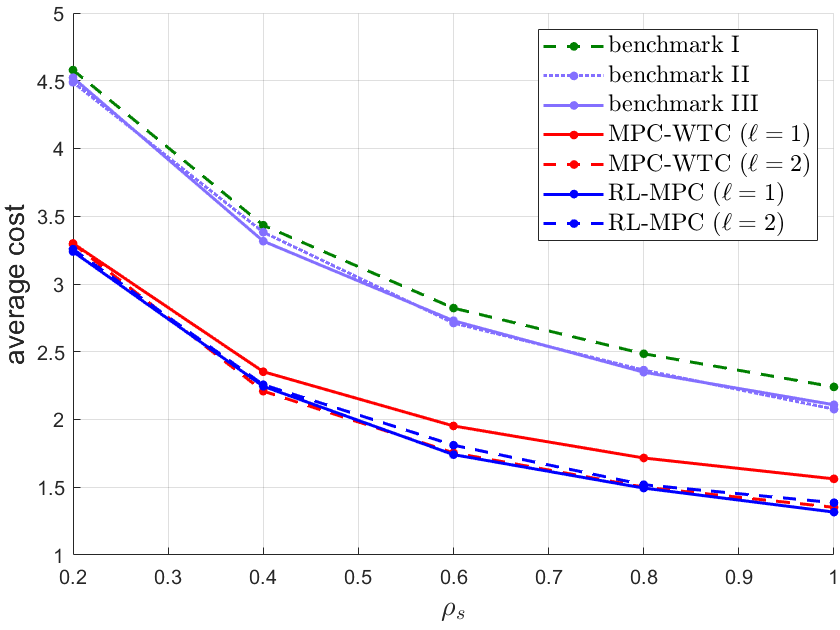} }
    \end{center}  
    \caption{Comparison of the studied policies for Scenario II in terms of the average cost for varying $\rho_s$ when a) $\lambda=0$, b) $\lambda=1$.}
    \label{fig:sim_rho}
\end{figure}

Fig.~\ref{fig:sim_rho} depicts the average cost as a function of the successful transmission probability $\rho_s$ for two values of the relative weight $\lambda$. We first observe that for lower values of $\rho_s$, benchmark III  outperforms benchmark II due to its erasure awareness, whereas the performance gap between these two policies diminishes with increasing $\rho_s$, as expected. On the other hand, the performances of both these round-robin policies lag behind benchmark I. Furthermore, we observe that MPC methods outperform all open-loop benchmark policies by utilizing the observations through belief, i.e., belief-dependent policies. We observe that choosing $\ell=1$ is sufficient for both MPC methods when $\lambda=0$. However, when $\lambda=1$, looking only at the next step in MPC-WTC ($\ell=1$) falls short compared to the other MPC policies.

\begin{figure}
    \centering
    \includegraphics[width=0.8\linewidth]{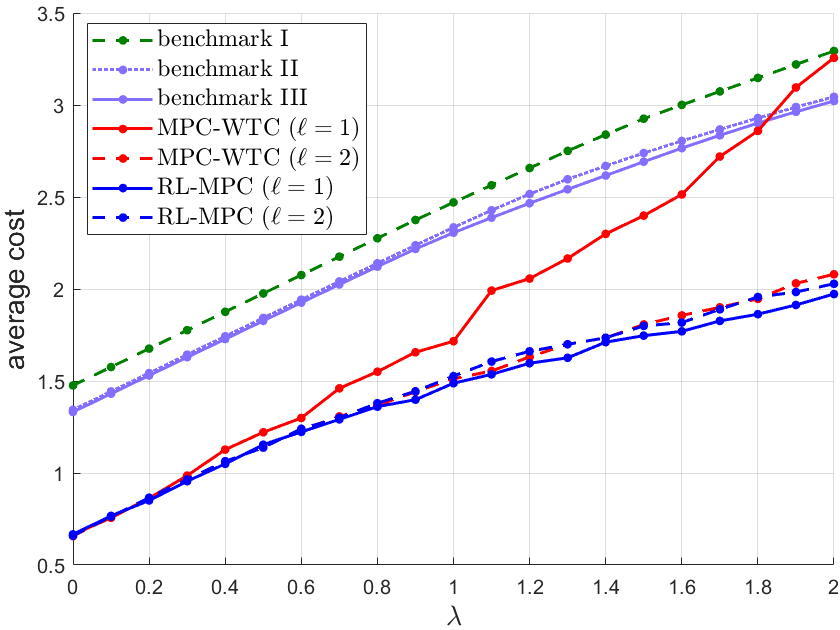}
    \caption{Comparison of the studied policies for Scenario II in terms of the average cost for varying values of $\lambda$ when $\rho_s=0.8$.}
    \label{fig:lam_2x3x2}
\end{figure}

In the second example, the average cost is depicted for varying values of $\lambda$ while fixing $\rho_s=0.8$, in Fig.~\ref{fig:lam_2x3x2}. As $\lambda$ increases, the monitor rarely sends pull requests, and the performance of MPC-WTC for $\ell=1$ degrades relative to the other belief-dependent MPC policies, which outperform the open-loop benchmark policies consistently.


\section{Conclusions} \label{sec:conc}
We investigated a remote estimation problem for which a monitor obtains different components of a joint Markov process through pull requests and the monitor aims to minimize a weighted combination of the age and the average sampling cost by answering two fundamental questions: when and which sensor to observe? First, we obtained sufficient statistics for solving the problem, namely the belief, and formulated the problem as a belief-MDP. Then, we proposed the use of model predictive control (MPC) to solve the belief-MDP with the MPC-WTC and RL-MPC belief-based policies, the latter policy also making use of reinforcement learning (RL) for finding the terminal costs off-line, whereas the main advantage of MPC-WTC is that it does not require any offline learning. 

We observed that both MPC methods outperformed three benchmark open-loop policies for all the examples we studied. We have also observed that RL-MPC policy performed very well for the look-ahead size $\ell=1$ and gains with increasing this parameter further to $\ell=2$ were limited. On the other hand, there were several examples for which MPC-WTC with $\ell=1$ performed poorly, whereas increasing the look-ahead size $\ell=2$ improved the performance substantially. Further performance gains were also observed to be limited for MPC-WTC, i.e., $\ell=3$ in the examples we studied. 

Consideration of other source dynamics other than the joint Markov structure, or the case when the source dynamics are not known, or only partially-known by the monitor, are left for future research. In addition, we assume that each time slot, the monitor can only send a pull request to at most a single sensor. However, the proposed methods are suitable for considering multiple pull requests simultaneously. 

\bibliographystyle{IEEEtran}
\bibliography{bibl}

\end{document}